\documentclass[aps,preprint,groupedaddress,showpacs,showkeys,floatfix]{revtex4}
\usepackage{amsmath}
\usepackage{amssymb}
\usepackage{graphicx}

\begin{document}

\title{The NNLO predictions for the rates of the 
\\ \boldmath{$W/Z$} production in 
\boldmath{$\stackrel{(-)~~}{pp}$} collisions\footnote{\it Contributed to 
21th International Symposium on Lepton and Photon 
Interactions at High Energies (LP 03), 11-16 Aug 2003, Batavia, Illinois}}
\author{S.~I.~Alekhin}

\affiliation{Institute for High Energy Physics, 142281 Protvino, Russia}

\begin{abstract}
{The NNLO rates of the intermediate vector boson production (IVB)
are calculated 
and found to agree with the preliminary results of Run II at the Fermilab 
$\overline{p}p$ collider.
The estimated uncertainties in NNLO predictions for 
the IVB rates including the errors in PDFs, $\alpha_{\rm s}$, and the 
factorization/renormalization scales are about 2\%
for the Fermilab collider and 3\% for the LHC
that allows to use these predictions 
as a competitive benchmark for calibration of the collision luminosity.}
\end{abstract}
\pacs{13.60.Hb,06.20.Jr,12.38.Bx}
\keywords{Intermediate vector bosons, hadron colliders}

\maketitle

The increase of the collision energy and intensity 
of the colliding beams in a new generation of the hadron colliders 
requires new approaches in the precise
monitoring of the collisions' luminosity
necessary for detecting manifestation of new physics.
The measurement of the rate of Intermediate Vector Bosons (IVB) production
is one of the promising tool for this purpose \cite{Dittmar:1997md}.
Due to large scale given by the IVB masses
the production cross section can be reliably calculated in the 
QCD-improved quark parton model. With the IVB masses and electroweak 
coupling well constraint from the wealth of other measurements
the largest source of the uncertainty in 
the calculated IVB rates comes from the high-order (HO) QCD corrections.
However the recent progress in the NNLO QCD calculations
allows to minimize the uncertainty due to missing HO corrections as well.
The NNLO coefficient functions for the Drell-Yan process
have been calculated~\cite{Hamberg:1990np}. 
Despite the NNLO anomalous dimensions 
are not known completely yet, the remaining uncertainty in  
the NNLO splitting functions~\cite{vanNeerven:2000wp} is at the level of 
several percents in the x-region relevant for existing data.
As a result the uncertainty in the NNLO PDFs due to incomplete knowledge
of the NNLO anomalous dimensions does not exceed the experimental 
uncertainties in the PDFs through the whole kinematics of the existing and 
planned hadron colliders~\cite{Alekhin:2001ih} and, therefore,  
calculations of the IVB rates up to the NNLO make sense.

\begin{figure}[h]
\includegraphics[width=14cm,height=12cm]{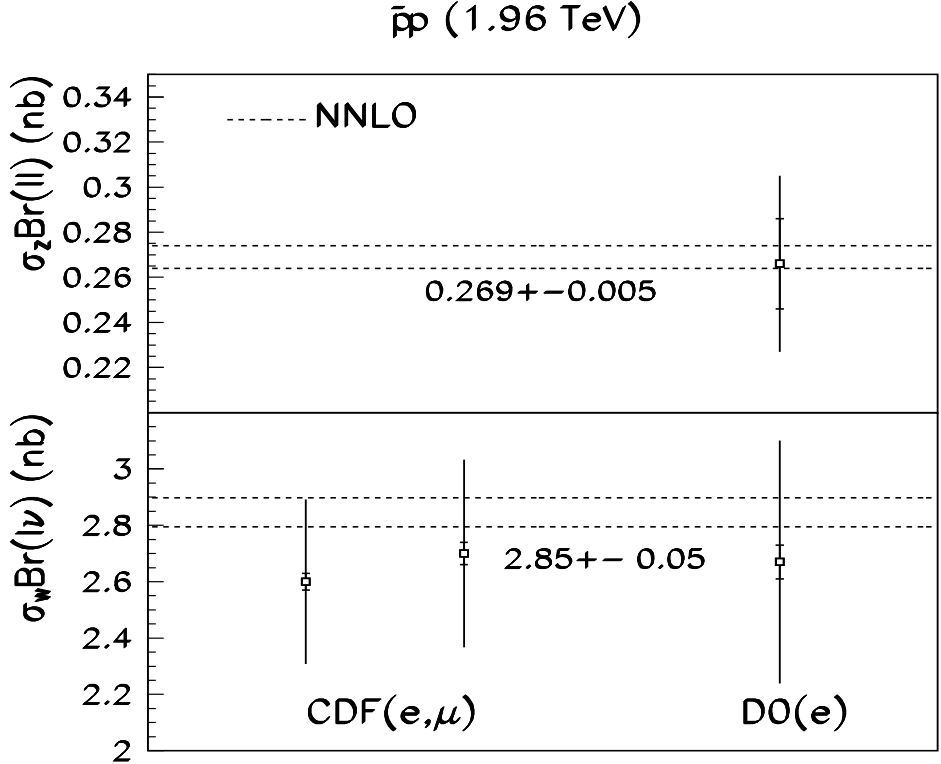}
\caption{The NNLO $W/Z$ production rates in 
the $\overline{p}p$ collisions at $\sqrt{s}=1.96~{\rm TeV}$ compared to the 
preliminary results for Run II. The area between dashes gives $1\sigma$ band 
uncertainty in the calculations. The inner bars of the data points give the 
statistical error, the outer ones give the
total errors including the luminosity uncertainty.}  
\label{fig:fnal}
\end{figure}

In this letter we give the NNLO calculations of the IVB rates 
at the Fermilab $\overline{p}p$ collider 
and the $pp$ Large Hadron Collider 
(LHC). The calculations are based on the 
code of Ref.\cite{Hamberg:1990np} with 
the NNLO PDFs of Ref.~\cite{Alekhin:2002fv} extracted from the 
fit to the global deep-inelastic-scattering (DIS) data. This choice of PDFs 
provides an advantage in comparison with the
Martin-Roberts-Stirling-Thorne (MRST) PDFs
of Ref.\cite{Martin:2002dr} since in the later case the PDFs are 
fitted using wider set of processes including the data on jet production 
for which the NNLO corrections are unknown. Besides, the MRST fit includes 
the data for IVB production as well and thus the predictions of the 
IVB rates based on these PDFs are not truly independent.

\begin{table}
\caption{\label{tab:fnal}
The production cross sections (in nb)
for the $W$-boson and $Z$-boson (in parenthesis)
in the $\overline{p}p$ collisions at $\sqrt{s}=1.96$~{\rm TeV}
calculated in different approximations for the PDFs and 
the coefficient functions.}
\begin{ruledtabular}
\begin{tabular}{ccc} 
PDFs  &     \multicolumn{2}{c}{Coefficient functions} \\ \cline {2-3} 

 &              NLO    &                NNLO \\

NLO         & 25.5(7.6)   &               26.2(7.8) \\

NNLO         & 25.9(7.7)    &              26.6(7.9) \\
\end{tabular}
\end{ruledtabular}
\end{table}
\begin{table}
\caption{\label{tab:lhc}
The same as Table~\protect\ref{tab:fnal} for the $pp$
collisions at $\sqrt{s}=14$~TeV.}
\begin{ruledtabular}
\begin{tabular}{ccc} 
PDFs  &     \multicolumn{2}{c}{Coefficient functions} \\ \cline {2-3} 

 &              NLO    &                NNLO \\

NLO   &       200.9(58.8)      &            200.6(58.8) \\

NNLO   &       204.4(59.9)    &              204.6(60.0) \\
\end{tabular}
\end{ruledtabular}
\end{table}

In our calculations the values of the IVB 
masses were set as $M_W=80.423~{\rm GeV}$, $M_Z=91.188~{\rm GeV}$,
the widths as $\Gamma_W=2.118~{\rm GeV}$, $\Gamma_Z=2.495~{\rm GeV}$,
the branching ratios of the IVB leptonic decays 
as $BR(W \rightarrow l\nu)=0.107$, $BR(Z \rightarrow ll)=0.034$,
squared sine of the Weinberg angle $x_W(M_{\rm Z})=0.2311$, 
squared cosine of the Cabibbo angle $c_C=0.9505$~\cite{Hagiwara:fs}.
The value of strong coupling constant 
$\alpha^{NNLO}_{\rm s}(M_{\rm Z})=0.1143$ used in the calculations
was found in the analysis of 
Ref.~\cite{Alekhin:2002fv} simultaneously with the parameterization of 
the PDFs. 

\begin{figure}[h]
\label{fig:rfnal}
\includegraphics[width=14cm,height=12cm]{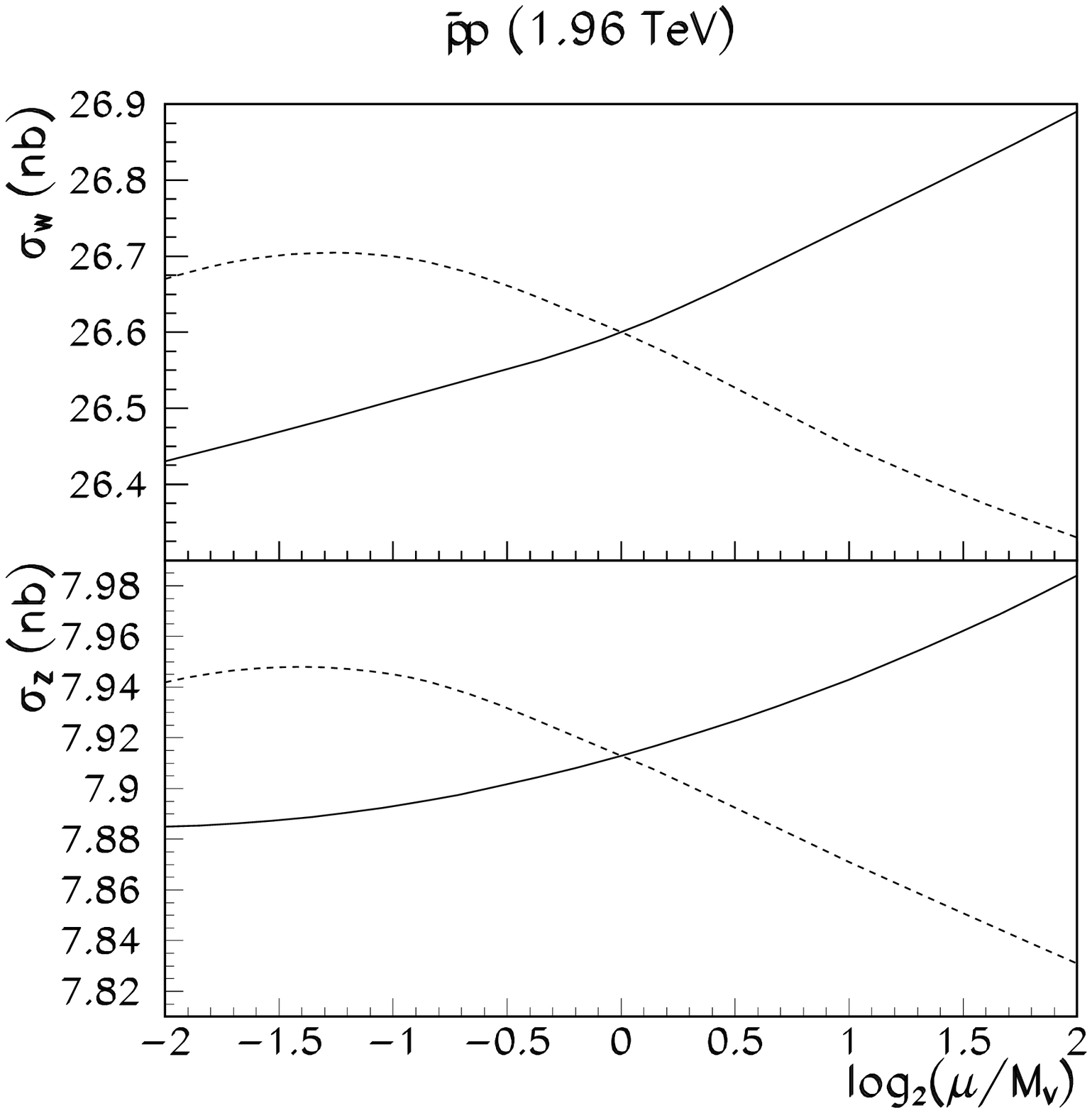}
\caption{Dependence of the NNLO predictions for the IVB rates
in the $\overline{p}p$ collisions at $\sqrt{s}=1.96$~{\rm TeV} 
on the factorization (solid lines) and the 
renormalization (dashes) scales $\mu$.}
\end{figure}
\begin{figure}[h]
\label{fig:rlhc}
\includegraphics[width=14cm,height=12cm]{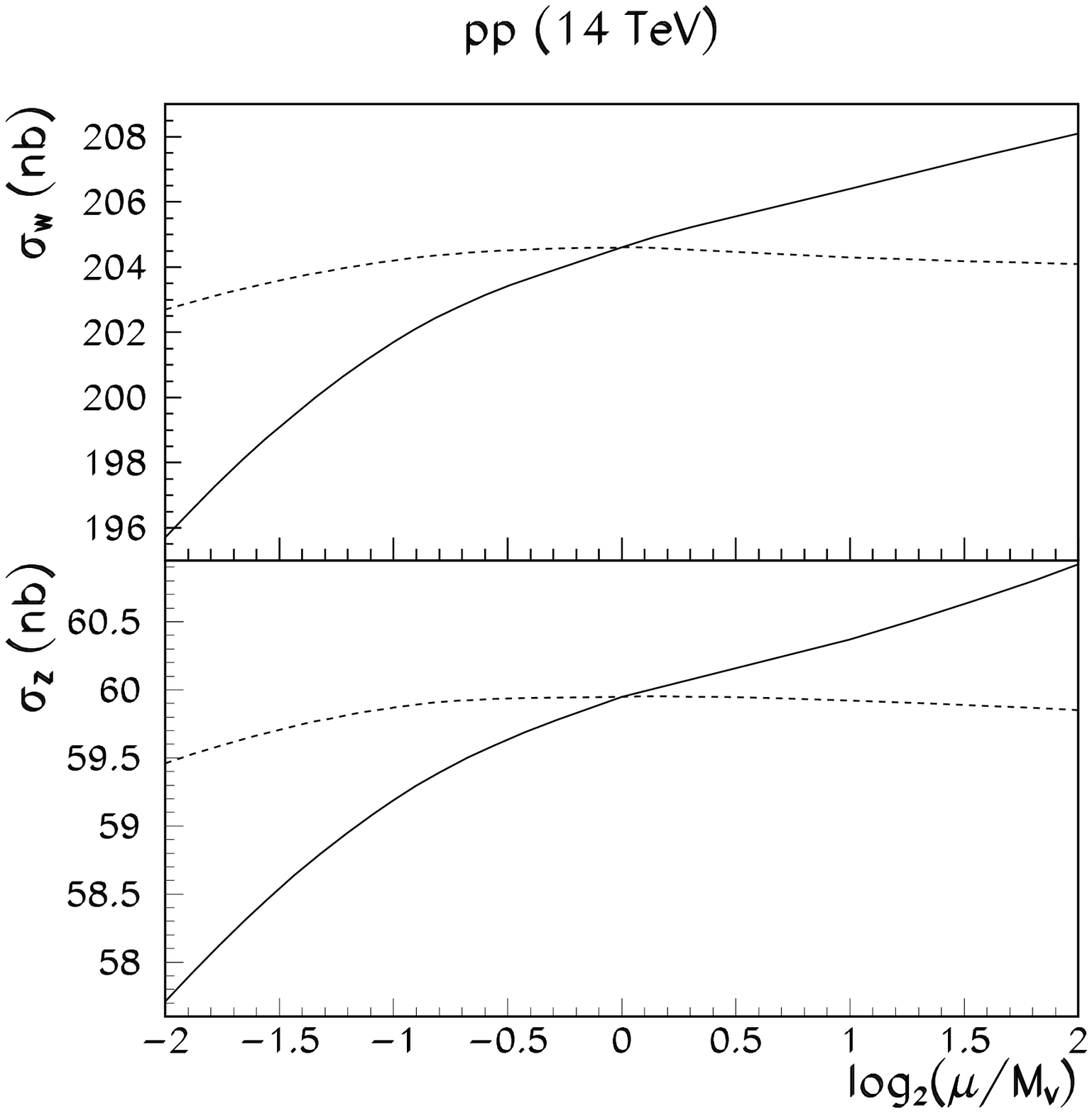}
\caption{The same as Fig.1 for the $pp$
collisions at $\sqrt{s}=14$~TeV.}
\end{figure}

The NNLO IVB production rates in the
$\overline{p}p$ collisions at $\sqrt{s}=1.96~{\rm TeV}$
calculated at the renormalization factorization scales set to  
the IVB masses are given in Fig.\ref{fig:fnal}.
The horizontal lines correspond to the $1\sigma$ bands of the predictions.
These bands stem from the PDFs errors and the error in $\alpha_{\rm s}$,
which is correlated with the PDFs errors
due to its value was extracted simultaneously with 
the PDFs from the same data set. The errors due to the high-twist (HT) 
contribution to the DIS cross sections
are also effectively included into the error bands given since 
the HT terms were also fitted in the analysis of Ref.~\cite{Alekhin:2002fv}. 
All errors were estimated by straightforward propagation of
the uncertainties in the inputs of the fit 
without introduction a scale factors.
This allows for rigorous treatment of 
these errors in terms of the probability theory and 
correct calculation of the confidence intervals.
The preliminary results of Run II for the IVB rates
\cite{Evans:2002wj} are given in Fig.\ref{fig:fnal} for comparison. 
The measured rates and predictions agree within the errors.
The errors in the measured rates due to the luminosity uncertainty are
typically about 10\% that is much larger than the errors in the 
predictions. This allows to use the predictions 
for the complementary cross-checks of the luminosity monitor.
For the $pp$ collisions at $\sqrt{s}=14~{\rm TeV}$
the NNLO IVB rates are estimated as 
$$
\sigma_{\rm Z}=60.0\pm1.9~{\rm nb},
$$
$$
\sigma_{\rm W}=204.6\pm6.4~{\rm nb}
$$
with the uncertainty  about 3\%.

To estimate the importance of the NNLO corrections on the IVB rates 
and to separate impacts of the NNLO corrections to 
the coefficient functions and to the PDFs we performed 
calculations combining the PDFs and the coefficient functions in different 
approximations. The value of $\alpha^{NLO}_{\rm s}(M_{\rm Z})=0.1171$ was 
used with the NLO PDFs,
in accordance with the results of the NLO fit of Ref.~\cite{Alekhin:2002fv}.
The results for the Fermilab collider and the LHC are given 
in Tables~\ref{tab:fnal},\ref{tab:lhc}.
For the Fermilab collider the both corrections have the same sign and 
comparable scale. For the LHC the effect of the NNLO corrections to the 
coefficient functions is marginal, while the change of the NLO PDFs by  
the NNLO ones causes sizeable increase of the cross sections 
with the relative scale of increase
comparable to the case of the Fermilab collider.
The latter is in disagreement with the results of
Ref.\cite{Martin:2000gq}, which
reported negative contribution of the NNLO corrections
to the IVB rates at the LHC. 
This disagreement can be attributed to the difference in the data sets used 
for fitting of the PDFs and needs further clarification.
Both for the Fermilab collider and the LHC the uncertainty in the IVB rates 
due to possible variation of the NNLO anomalous dimensions
is less than 1\%.
 
An additional source of the uncertainty in the 
predictions for the IVB rates is variation of the 
factorization and the renormalization scales. 
However, in the NNLO these uncertainties 
are greatly suppressed as compared to the NLO case~\cite{Hamberg:1990np}.
As one can see in Figs.2,3
the factorization uncertainty in the NNLO IVB rates 
estimated for very wide variation of the scale
is less than 1\% for the Fermilab collider and $2\div3\%$
for the LHC, while the errors due to the renormalization scale 
are generally smaller
than ones due to the factorization scale. In conclusion, 
the estimated uncertainties in the NNLO predictions for 
the IVB rates including the errors in PDFs, $\alpha_{\rm s}$, and the 
factorization/renormalization scales are about 2\%
for the Fermilab collider and 3\% for the LHC
that allows to use these predictions 
as a competitive benchmark for calibration of the collision luminosity. 

I am indebted to S.~Kulagin for reading the manuscript and valuable comments,
W.~Giele and C.P.~Yuan for stimulating discussions.
The work was supported by the RFBR grant 03-02-17177.

\end{document}